# Attractive force on atoms due to blackbody radiation

Philipp Haslinger[1], Matt Jaffe[1], Victoria Xu[1], Osip Schwartz[1,2], Matthias Sonnleitner[3], Monika Ritsch-Marte[4], Helmut Ritsch[5], and Holger Müller[1,2,*]

[1]Department of Physics, University of California–Berkeley, Berkeley, CA, USA.
[2]Molecular Biophysics and Integrated Bioimaging, Lawrence Berkeley National Laboratory, CA, USA.
[3]School of Physics and Astronomy, University of Glasgow, United Kingdom
[4]Division for Biomedical Physics, Medical University of Innsbruck, Innsbruck, Austria
[5]Institute for Theoretical Physics, University of Innsbruck, Innsbruck, Austria

*Correspondence to: hm@berkeley.edu

**Objects at finite temperature emit thermal radiation with an outward energy-momentum flow, which exerts an outward radiation pressure. At room temperature, a cesium atom scatters on average less than one of these blackbody radiation photons every $10^8$ years. Thus, it is generally assumed that any scattering force exerted on atoms by such radiation is negligible. However, atoms also interact coherently with the thermal electromagnetic field. In this work, we measure an attractive force induced by blackbody radiation between a cesium atom and a heated, centimeter-sized cylinder which is orders of magnitude stronger than the outward directed radiation pressure. Using atom interferometry, we find that this force scales with the fourth power of the cylinder`s temperature. The force is in good agreement with that predicted from an ac Stark shift gradient of the atomic ground state in the thermal radiation field[1]. This observed force dominates over both gravity and radiation pressure, and does so for a large temperature range.**

Quantum technology continues to turn formerly unmeasurable effects into technologically important physics. For example, minuscule shifts of atomic energy levels due to room-temperature blackbody radiation have become leading influences in atomic clocks at or beyond the $10^{-14}$ level of accuracy [2]. For this reason, they have become relevant to, e.g., timekeeping, navigation, and geodesy. Thermal radiation from a heated source should also result in a repulsive radiation pressure on atoms through absorption of photons[3–6]. However, the photon energies around room temperature are far off-resonance, thus the scattering rate is low and scattering induced force very small, leading to mm/s velocity changes in hundreds of thousands of years for, e.g., the cesium D line. Here, we show that such blackbody radiation produces a much larger acceleration at the μm/s$^2$-level pointing towards the source, even near room temperature. It is well-described by the intensity gradient of blackbody radiation that gives rise to a spatially-dependent ac Stark shift[1], similar to the dipole forces induced by lasers in optical tweezers[7], atom trapping[8], coherent manipulation of atoms[9], or of molecular clusters[10]. We expect it to be the dominant force on polarizable objects over a large temperature range[1] and thus important in atom interferometry, nanomechanics, or optomechanics[11]. Controlling this force will enable higher precision in atom interferometers including: tests of fundamental physics such as of the equivalence principle[12–14], planned searches for dark matter and dark energy[15], gravity gradiometry[16], inertial navigation and perhaps even Casimir force measurements and gravitational wave detection[17,18].

As shown in Fig. 1, we perform atom interferometry with cesium atoms[19] in an optical cavity to measure the force induced by blackbody radiation. Our setup is similar to the one we used previously[20,21]. Cesium atoms act as matter waves in our experiment. They are laser-cooled to a temperature of about 300 nK and launched upwards into free fall. We then manipulate them with counterpropagating laser beams, which "kick" the atoms with an impulse $\hbar k_{\text{eff}}$ from two photons. The intensity and the duration of the laser pulses determine whether we transfer the atom with a 50 % probability (a "π/2 -pulse") or nearly 100 % (a "π-pulse"), respectively. We apply a π/2-π-π/2 pulse sequence, spaced by intervals of $T$ = 65 ms, that splits, redirects, and recombines the free falling atomic wavefunction, forming a Mach-Zehnder atom interferometer. The matter waves propagate along the two interferometer arms while accumulating an acceleration phase difference $\Delta\phi = k_{\text{eff}} a_{\text{tot}} T^2$, where $a_{\text{tot}}$ is the total acceleration experienced by the atom in the lab frame. The probability of the atom to exit the interferometer in one of the outputs is given by $P = \cos^2(\Delta\phi/2)$.

For the heated object, we use a non-magnetic metal (tungsten) cylinder of 25.4 mm height and diameter. The laser beam passes the cylinder through a 10-mm bore at its center. The cylinder also has a 5-mm slit on the side, which allows us to change its position between a location

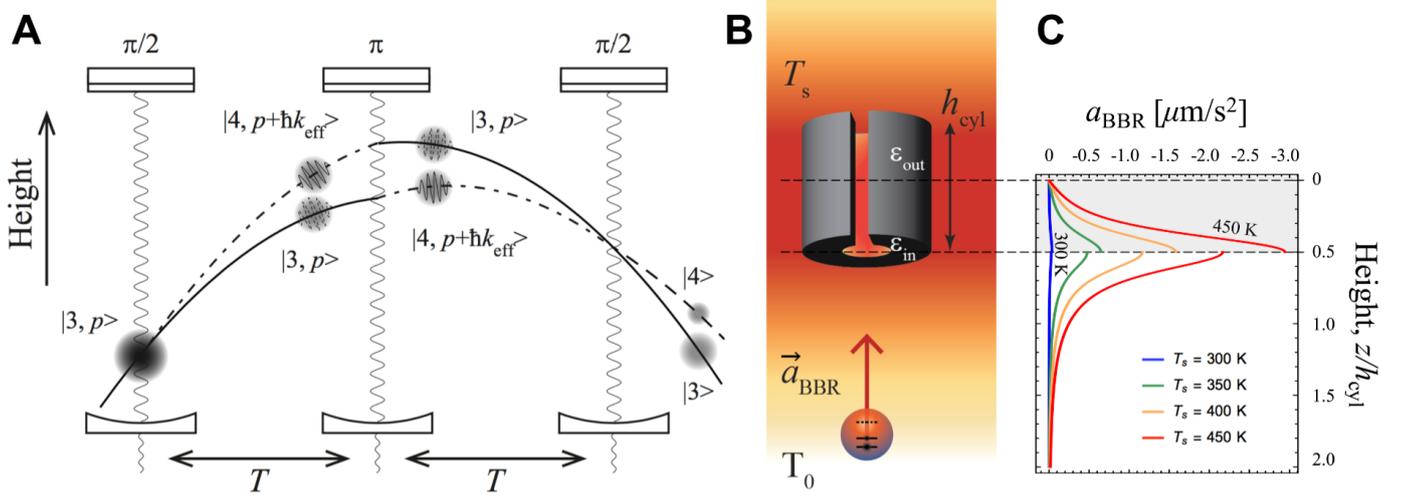

**Figure 1 | Setup.** (A) Space-time diagram of each atom's trajectories in our Mach-Zehnder interferometer. (B) The intensity gradient of blackbody radiation surrounding a heated, hollow cylinder causes a force on atoms. (C) Theoretical calculation of the acceleration of cesium atoms due to blackbody radiation, $a_{BBR}$, as a function of the distance $z$ along the cylindrical axis. The vertical axis is taken from the center of the source mass. The gray shaded area marks the region inside the hollow core of the cylinder. Discontinuities in the theoretical acceleration curve stem from edge effects of the hollow cylinder.

close to the atom interferometer and a remote one[21], without interrupting the cavity mode.

At the start of each experimental run, we heat the cylinder to a temperature of about 460 K with an infrared laser, which is subsequently switched off. We then measure the acceleration of the atoms during the cool-down period of up to 6 hours, while we monitor the temperature with an infrared sensor. When the source mass has cooled to near room temperature, we re-heat it to start another run. Acceleration data was acquired during 63 "cooldowns". Changing the cylinder´s distance to the atom interferometer allows us to separate source-mass induced forces from other forces, in particular the million-fold larger one from Earth's gravity. The near position exposes the atoms to blackbody radiation arising from the source, while the far position serves as a reference. We then investigate the temperature dependence of the acceleration difference.

Fig. 2 shows this measured acceleration as a function of the source mass temperature. We fit the data with a power-law $a_{BBR} = C\,(T_s^n - T_0^4)$, where $a_{BBR}$ is the acceleration difference measured between the near and far positions of the cylinder, $T_s$ the source temperature, $n$ an exponent, $T_0$=296 K the temperature of the environment, and $C$ a factor of proportionality that will be related to the albedo and geometry of the source. The fit parameters are $C$ and $n$. We obtain an exponent of $n= 4.021 \pm 0.035$, compatible with the $T_s^4$ dependence expected from effects of blackbody radiation.

It is important to rule out artifacts which could partially mimic a blackbody-induced acceleration. For example, spatially constant energy-level shifts induced by the blackbody radiation (rather than an ac Stark shift gradient, which produces a force) can be ruled out for multiple reasons (see supplement). For example, they would be common to both interferometer arms, and thus cancel out. The pressure applied by hot background atoms from outgassing of the heated source mass removes a substantial fraction of the cold atoms from the detection region at its highest temperatures, so it is conceivably a component of the measured force on the remaining atoms. This, however, can be ruled out by multiple observations. First, this pressure should push the atoms away from the source, while the observed acceleration is towards the source. Second, it should depend exponentially on the source mass temperature; such an exponential component is not evident in the data. Finally, any scattering of hot background atoms with atoms that take part in the interferometer would be incoherent, and would reduce the visibility of our interference fringes. Fig. 3, however, shows that the visibility is constant over our temperature range, ruling out scattering. This observation also confirms that absorption or stimulated emission of incoherent blackbody photons is negligible (see Fig.4). Casimir forces[22–24] are irrelevant since the atoms never come closer to source mass surface than about 2 mm.

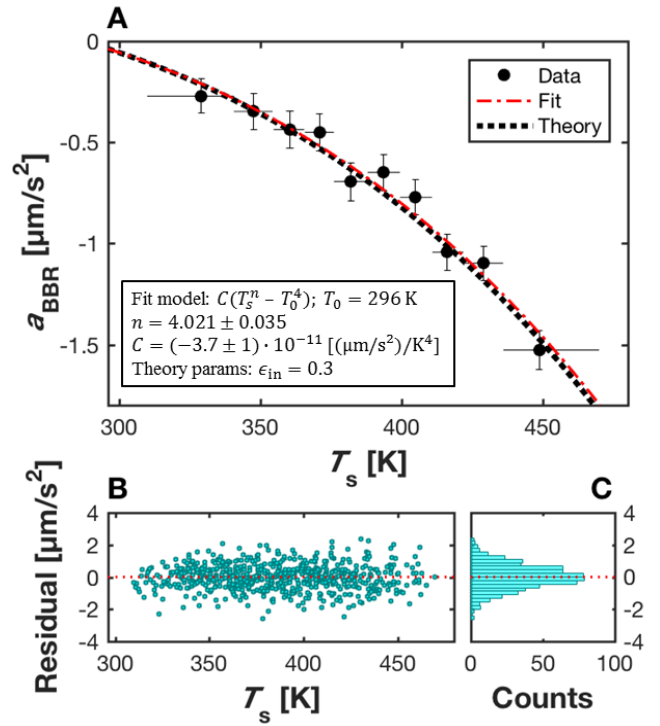

**Figure 2 | Experimental data.** Measured acceleration as a function of the source mass temperature $T_s$. A quartic dependence on $T_s$ is observed for the acceleration experienced by cesium atoms towards the source mass. (A) Data from 63 thermal cycles are binned in temperature with $N_{bin}$=65 measurements per bin. The black dots represent the weighted mean of each bin. Vertical error bars show 1-sigma uncertainty on the weighted mean. Horizontal bars show the temperature spread of the $N_{bin}$ measurements in the bin. The red dot-dash line is the fit. The black dotted line represents a theoretical calculation of the impulse imparted to the atoms during interferometry. (B) Residuals from the bulk acceleration data (cyan) to the quartic fit (red dot-dash line), and a histogram of the bulk residuals (C) exhibit a Gaussian distribution around 0.

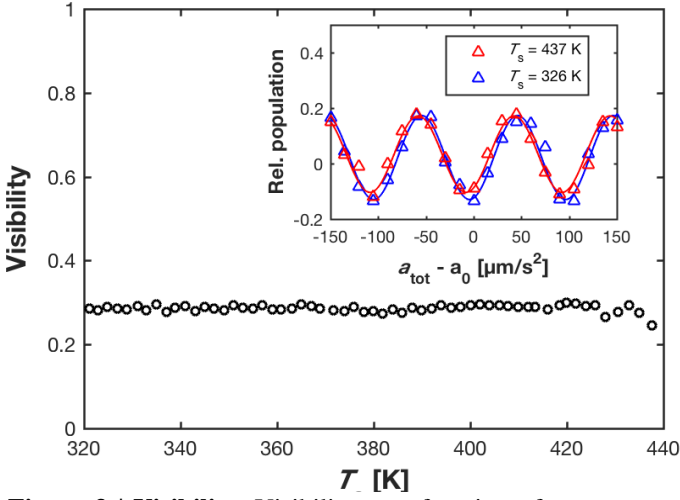

**Figure 3 | Visibility.** Visibility as a function of temperature, averaged in bins of 2 Kelvin for clarity. Scattering or absorption of photons would lead to a dephasing of the atomic ensemble, resulting to a reduction of visibility. No obvious loss of visibility is a strong indication that the contribution of scattering and absorption events is negligible.

We now explain the measured acceleration in terms of a force due to the gradient in the ground-state energy level shift (ac Stark effect) induced by blackbody radiation, despite this energy-level shift being only $h \times 15$ Hz at our highest temperatures, where $h$ is the Planck constant. For the relevant temperature range, nearly all thermal radiation has a frequency well below the cesium D-line, so that radiation pressure from absorption and emission of blackbody radiation is negligible. The shift of the atomic ground state energy can be approximated by using the atom's dc polarizability[25] $\alpha_{Cs} \approx h \times 0.099$ Hz/(V/cm)$^2$ as[2] $\Delta E(r) = -\alpha_{Cs} u(r)/(2\varepsilon_0)$, where $u(r)$ is the electromagnetic energy density for the thermal field measured at a distance $r$ from the source, and $\varepsilon_0$ is the vacuum permittivity. For isotropic blackbody radiation at the temperature $T_s$ of the source, we have $u = 4\sigma T^4/c$ (where $c$ is the speed of light), and

$$\Delta E_0 = -2\frac{\alpha_{Cs}\, \sigma T_s^4}{c\varepsilon_0} \qquad (1)$$

where $\sigma$ is the Stefan-Boltzmann constant. If the heated body is a sphere of radius $R$, then the sphere's blackbody radiation will dilute with distance, with energy density $u(r)$ proportional to $R^2/(4r^2)$. Taking the gradient gives the acceleration from the blackbody radiation force[1] in spherical geometry:

$$a = -\alpha_{Cs} \frac{\sigma(T_s^4 - T_0^4)}{c\,\varepsilon_0\, m_{Cs}} \frac{R^2}{r^3} \qquad (2)$$

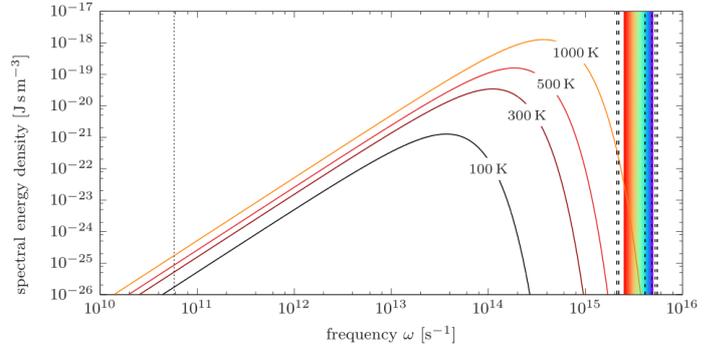

**Figure 4 | Blackbody radiation.** The spectra of blackbody radiation for various temperatures compared to transition frequencies of ground-state cesium indicated by vertical lines. The dash-dotted line at the left refers to the hyperfine splitting of the ground state used in the current definition of the second. The dashed lines at the right are strong absorption lines starting from the D$_1$ transition at $\omega \approx 2\pi \cdot 335$ THz. The colorful band indicates the visible spectrum as a guide for the eye.

The force points radially inwards.

For a more detailed calculation, we model the tungsten cylinder as an opaque diffuse-gray surface whose absorptivity $\alpha$ and emissivity $\epsilon = \alpha$ are independent of direction[26] and whose reflectivity $\rho = 1-\epsilon$ is constant over the considered temperature range. We have measured the cylinder's emissivity at the bottom surface facing the atom to be $\epsilon = 0.35 \pm 0.05$ by using an infrared temperature sensor. The radiation experienced by the atom is the sum of the thermal fields coming from the source mass surface of temperature $T_s$ and the ambient radiation inside the vacuum chamber at temperature $T_0$. Our model accounts for ambient radiation reflected by the outer surface of the cylinder, and for the fact that radiation can make multiple reflections inside the bore of the cylinder. This effectively increases the emissivity from that region[26]. The model thus predicts the radiation intensity, and therefore the resulting forces on the cesium atoms as a function of position, as shown in Fig. 1. The dashed line in Fig. 2 shows the predicted net effect for a source mass emissivity of $\epsilon = 0.3$, within the range of measured emissivities. The agreement is apparent.

Just as blackbody radiation affects atomic clocks[2,27], the acceleration due to the blackbody field gradient observed here influences any high precision acceleration measurements with polarizable matter, including atomic and molecular interferometers, experiments with nanospheres, and potentially measurements of the Casimir effect and gravitational wave detectors. For example, inside a thin cylindrical vacuum chamber, the thermal radiation field nearly follows the local temperature $T(z)$ of the walls, inducing an acceleration of atoms of

$$a(z) = \frac{1}{m_{At}} \frac{\partial}{\partial z} \frac{2\alpha_{At}\, \sigma\, T^4(z)}{c\varepsilon_0} \qquad (3)$$

where $m_{At}$ and $\alpha_{At}$ are the atom's mass and static polarizability. Simulations confirm this approximation for thin cylinders, even for walls with percent-level emissivity. For cesium atoms, e.g., a linear temperature gradient of $T' = 0.1$ K/m around a base of 300 K would result in $a \approx 10^{-11}$ m/s$^2$, non-negligible in, e.g., terrestrial and space-borne high precision measurements including tests of the equivalence principle, gravity gradiometers or gravitational wave detection with atom interferometry. Effects will be suppressed in nearly overlapped simultaneous conjugate interferometers[28] used for measuring the fine structure constant[29,30]. The acceleration can be mitigated by monitoring and/or equalizing the temperature across the vacuum chamber or (as shown by our simulations) by using wide, highly reflective vacuum chambers, wherein multiple reflections make the thermal radiation more isotropic. On the other hand, blackbody radiation can be used to simulate potentials. For example, heated test masses could be used to calibrate an atom interferometer for measuring the gravitational Aharonov-Bohm effect[31].

We thank Randy Putnam for collaboration in the lab and Dennis Rätzel for stimulating discussions. This material is based upon work supported by the David and Lucile Packard Foundation, the National Science Foundation under grant No. 037166, the Defense Advanced Research Projects Agency grant No. 033504, as well as the National Aeronautics and Space Administration Grants No. 041060-002, 041542, 039088, 038706, and 036803. We also acknowledge collaboration with Honeywell Aerospace under DARPA Contract No. N66001-12-1-4232. O. S. was supported by HFSP fellowship LT000844/2016-C. M. S. was supported by the ERC AdG(247024 catchIT). P. H. and M. S. thank the Austrian Science Fund (FWF): J3680, J3703.

**References:**

1. Sonnleitner, M., Ritsch-Marte, M. & Ritsch, H. Attractive Optical Forces from Blackbody Radiation. *Phys. Rev. Lett.* **111,** 23601 (2013).
2. Safronova, M. S. *et al.* Black-body radiation shifts and theoretical contributions to atomic clock research. *IEEE Trans. Ultrason. Ferroelectr. Freq. Control* **57,** 94–105 (2010).
3. Nichols, E. & Hull, G. Pressure due to light and heat radiation. *Astrophys. J.* **15,** 62 (1902).
4. Lebedev, P. N. & Lazarev, P. P. *Die Druckkräfte des Lichtes, zwei Abhandlungen*. **188,** (W. Engelmann, 1913).
5. Burns, J. A., Lamy, P. L. & Soter, S. Radiation forces on small particles in the solar system. *Icarus* **40,** 1–48 (1979).
6. Shestakova, L. I. Solar radiation pressure as a mechanism of acceleration of atoms and first ions with low ionization potentials. *Sol. Syst. Res.* **49,** 139–145 (2015).
7. Ashkin, A. & Dziedzic, J. M. Optical trapping and manipulation of viruses and bacteria. *Science.* **235,** 1517–1520 (1987).
8. Phillips, W. D. Nobel Lecture: Laser cooling and trapping of neutral atoms. *Rev. Mod. Phys.* **70,** 721–741 (1998).
9. Cronin, A. D., Schmiedmayer, J. & Pritchard, D. E. Optics and interferometry with atoms and molecules. *Rev. Mod. Phys.* **81,** 1051–1129 (2009).
10. Hornberger, K., Gerlich, S., Haslinger, P., Nimmrichter, S. & Arndt, M. Colloquium: Quantum interference of clusters and molecules. *Rev. Mod. Phys.* **84,** 157–173 (2012).
11. Aspelmeyer, M., Kippenberg, T. J. & Marquardt, F. Cavity optomechanics. *Rev. Mod. Phys.* **86,** 1391–1452 (2014).
12. Dimopoulos, S., Graham, P., Hogan, J. & Kasevich, M. Testing General Relativity with Atom Interferometry. *Phys. Rev. Lett.* **98,** 111102 (2007).
13. Schlippert, D. *et al.* Ground Tests of Einstein's Equivalence Principle: From Lab-based to 10-m Atomic Fountains. *arXiv: 1507.05820* (2015).
14. Tino, G. M. *et al.* Precision Gravity Tests with Atom Interferometry in Space. *Nucl. Phys. B - Proc. Suppl.* **243–244,** 203–217 (2013).
15. Hamilton, P. *et al.* Atom-interferometry constraints on dark energy. *Science.* **349,** 849–851 (2015).
16. McGuirk, J. M., Foster, G. T., Fixler, J. B., Snadden, M. J. & Kasevich, M. A. Sensitive absolute-gravity gradiometry using atom interferometry. *Phys. Rev. A* **65,** 33608 (2002).
17. Canuel, B. *et al.* Exploring gravity with the MIGA large scale atom interferometer. *arXiv:1703.02490* (2017).
18. Graham, P. W., Hogan, J. M., Kasevich, M. A. & Rajendran, S. New method for gravitational wave detection with atomic sensors. *Phys. Rev. Lett.* **110,** (2013).
19. Peters, A., Chung, K. & Chu, S. Measurement of gravitational acceleration by dropping atoms. *Nature* **400,** 849–852 (1999).
20. Hamilton, P. *et al.* Atom Interferometry in an Optical Cavity. *Phys. Rev. Lett.* **114,** 100405 (2015).
21. Jaffe, M. *et al.* Testing sub-gravitational forces on atoms from a miniature, in-vacuum source mass. *Nat. Phys.* (2017). doi:10.1038/nphys4189
22. Bordag, M., Mohideen, U. & Mostepanenko, V. M. New developments in the Casimir effect. *Phys. Rep.*

**353,** 1–205 (2001).
23. Lamoreaux, S. K. The Casimir force: background, experiments, and applications. *Reports Prog. Phys.* **68,** 201–236 (2005).
24. Scheel, S. & Buhmann, S. Y. Casimir-Polder forces on moving atoms. *Phys. Rev. A* **80,** 42902 (2009).
25. Gregoire, M. D., Brooks, N., Trubko, R. & Cronin, A. D. Analysis of polarizability measurements made with atom interferometry. *Atoms* **4,** 21 (2016).
26. Robert, J. R. H. *Thermal Radiation Heat Transfer FIFTH EDITION*. (CRC Press, 2015).
27. Nicholson, T. L. *et al.* Systematic evaluation of an atomic clock at $2 \times 10^{-18}$ total uncertainty. *Nat. Commun.* **6,** 6896 (2015).
28. Chiow, S., Herrmann, S., Chu, S. & Müller, H. Noise-Immune Conjugate Large-Area Atom Interferometers. *Phys. Rev. Lett.* **103,** 50402 (2009).
29. Bouchendira, R., Cladé, P., Guellati-Khélifa, S., Nez, F. & Biraben, F. New Determination of the Fine Structure Constant and Test of the Quantum Electrodynamics. *Phys. Rev. Lett.* **106,** 80801 (2011).
30. Parker, R. H. *et al.* Controlling the multiport nature of Bragg diffraction in atom interferometry. *Phys. Rev. A* **94,** 53618 (2016).
31. Hohensee, M., Estey, B., Hamilton, P., Zeilinger, A. & Müller, H. Force-Free Gravitational Redshift: Proposed Gravitational Aharonov-Bohm Experiment. *Phys. Rev. Lett.* **108,** 230404 (2012).

## Methods:

**Atom interferometer.** Cesium atoms are magneto-optically trapped inside an ultra-high vacuum chamber, laser-cooled to a temperature of about 300 nK using Raman sideband cooling[32] and prepared in the magnetically insensitive $F = 3$, $m_F = 0$ hyperfine ground state. We use laser pulses enhanced by the optical cavity to manipulate the atomic wavepackets (Fig. 1A). An atom in the $F = 3$ state with momentum $p_0$ absorbs a photon with momentum $+\hbar k$ and is stimulated to emit a photon with momentum $-\hbar k$. The atom emerges in the $F=4$ state and at a momentum of $p_0+\hbar k_{eff}$, where $k_{eff}=2k$. We can set the intensity and the duration of the laser pulses to transfer the atom with a 50 % probability (a "π/2 -pulse") or nearly 100 % (a "π-pulse"), respectively. A π/2-π-π/2 pulse sequence with pulses separated by a time $T = 65$ ms splits, redirects and recombines the free falling atomic wavefunction, forming a Mach-Zehnder interferometer. Along the trajectory, the two interferometer arms accumulate an acceleration phase difference $\Delta\phi = k_{eff}\, a_{tot}\, T^2$, where $a_{tot}$ is the acceleration experienced by the atom in the lab frame. The probability of the atom to exit the interferometer in state $F=3$ is given by $P=\cos^2(\Delta\phi/2)$. Since the atoms are in free fall under the earth's gravity, we chirp the laser frequencies in the laboratory frame at a rate of $\simeq 23$ MHz/s, so that the laser beams stay on resonance in the atoms' frame of reference.

For efficient detection of the $\sim 10^5$ atoms at the interferometer output, we reverse the launch sequence to catch the sample. Non-participating atoms that have left the cavity mode due to thermal motion, fall away. A pushing beam separates the state-labeled outputs of the interferometer. They are counted by fluorescence detection to determine $P$.

A single acceleration measurement is taken by adjusting the rate of the gravity-compensation chirp to trace out oscillations of $P$ with $\Delta\phi$. Fitting this fringe to a sine wave allows to extract the phase, and thus the acceleration experienced by the atoms. Eight fringes are taken consecutively before toggling the source mass position.

**Test mass.** The heated object is suspended inside the vacuum chamber by a non-magnetic (titanium) threaded rod (2.5 mm diameter) with a relatively low thermal conductivity of about 2 mW/K. We heat the cylinder by shining a Nd:YAG fiber laser (IPG Photonics YLR-100-1064LP) through the slit into the bore, where it is better absorbed due to multiple reflections. Within 12 min and a laser power of 8 W, we heat the cylinder from room temperature to about 460 K.

**Outgassing of the source mass.** The background pressure varies with source mass temperature. Initially, outgassing of the cylinder at 460 K caused a pressure increase to $\sim 10^{-7}$ mbar from a room temperature vacuum of $\sim 10^{-10}$ mbar (measured by an ion gauge about 50 cm away from the cylinder). After several heating cycles, this pressure increase was reduced to $\sim 10^{-9}$ mbar.

**Temperature measurement.** The temperature is measured using an infrared temperature sensor (Omega OS150 USB2.2, spectral response: 2.0 to 2.4 µm) through the vacuum chamber windows, which are made of fused silica and have a transmission cutoff just under $\lambda \approx 3$ µm. The infrared sensor works across a temperature range of 320 - 440 K; outside of this range, we can determine the temperature of the cylinder by extrapolation. This extrapolation is performed by calibrating the cooling curves to a heat-loss differential equation including both conduction and radiation.

**Systematic effects.** Possible artifacts which could influence this observation are well understood and can be ruled out:

*Constant Stark shifts* – In addition to the cancellation between interferometer arms mentioned in the main text, spatially constant ac Stark shifts would also be common to both ground-state hyperfine states, and thus cancel out even within each interferometer arm. This is because the blackbody radiation is very far detuned from any optical transition in the atom, and thus causes the same energy level shift to both hyperfine ground states. To verify, we performed the interferometer with opposite sign wave vector $\pm\mathbf{k}_{eff}$, implementing so-called "k-reversal"[16]. This inverts the signal $\mathbf{k}_{eff} \cdot \mathbf{a}_{tot}T^2$ arising from acceleration $\mathbf{a}_{tot}$ but would not invert a simple ac Stark phase. We observe that the effect inverts sign with $\mathbf{k}_{eff}$, as expected for a force. Our results in Fig. 2 include data runs for both directions of the wavevector, performed independently, confirming a real acceleration.

*Magnetic fields* – The magnetic fields are identical to those in [21]. Phase shifts due to source dependent magnetic fields give rise to an acceleration of only $- 2.5 \pm 11$ nm/s², less than 1% of the blackbody induced acceleration.

*Thermal expansion* - Heating of the cylinder eventually transfers heat to the vacuum chamber, potentially causing thermal expansion. This could affect the interferometer by, e.g., changing the cavity length. Such thermal expansion is avoided using a slow temperature feedback loop to hold the cavity distance constant throughout the experiment.

*Surface effects* - Casimir forces[22–24] are suppressed since the atoms never come closer to source mass surface than about 2 mm.

*Other effects* – A more comprehensive analysis of systematic effects was carried out in[21] using the same experimental setup. All effects analyzed are found to be below percent level compared to the blackbody force.

**Modeling.** The inner surface of the cylinder was not accessible with the IR temperature sensor due to geometrical constrains. However, we assume similar emissivities due to similar surface finishes. The radiation experienced by the atom is the sum of the thermal fields coming from the source mass surface of temperature $T_s$ and the ambient radiation inside the vacuum chamber at temperature $T_0$. From the atom's position $z$ each of the $i=1,…,N$ radiating and reflecting surfaces covers a solid angle $\Omega_i(z)$ such that the total shift of the ground state energy level is given by

$$\Delta E(z) = -\sum_i \frac{\Omega_i(z)}{4\pi} \frac{\alpha_{Cs}}{2} \frac{4}{\varepsilon_0 c} J_i$$

Where $J_i$ denotes the radiant energy per unit area (radiocity) coming from the $i^{th}$ surface; for a black surface that is $J_i=\sigma T_i^4$. For a diffuse gray body of emissivity $0<\epsilon_i<1$, this changes to $J_i=\epsilon_i \sigma T_i^4+(1-\epsilon_i)G_i$, where $G_i$ is the radiation flux coming towards that surface, which is then reflected towards the atom.

The outer surface of the cylinder reflects some of the ambient radiation such that $J_{out}= \epsilon\sigma T_s^4 +(1-\epsilon)\sigma T_0^4$. For the inner surface of the cylinder we account for internal reflection which effectively increases the emissivity from that region[26]. The vacuum chamber itself is assumed large enough such that we can ignore radiation coming from the cylinder and reflected by the walls of the vacuum chamber back to the atom. Finally, we also ignored that a segment has been cut out of the probe, see Fig. 1, and assume a radially symmetric hollow-core cylinder. Combining all these considerations we can calculate the spatial dependence of the blackbody radiation intensity and therefore, the level shift and the resulting forces on the cesium atoms as they approach the cylinder, as shown in Fig. 1. The jump in the acceleration at $z=h/2$ is a result of the sudden change in geometry, seen by the atom, as it enters the hollow cylinder. As the cylinder is cut open on one side this change will not be as pronounced for the actual setup.

**References:**

32. P. Treutlein, K. Y. Chung, S. Chu, High-brightness atom source for atomic fountains. *Phys. Rev. A.* **63**, 51401 (2001).